\documentclass{nature}
\usepackage{amssymb,amsmath}
\usepackage{CJK}
\usepackage{graphicx,color}
\usepackage{lipsum}
\usepackage{hyperref}
\usepackage{lineno}

\usepackage{stackrel}
\makeatletter
\let\saved@includegraphics\includegraphics
\AtBeginDocument{\let\includegraphics\saved@includegraphics}
\renewenvironment*{figure}{\@float{figure}}{\end@float}
\makeatother

\begin{document}
\title{Contact force measurements and local anisotropy in ellipses and disks}

\author{Yinqiao Wang$^{1,2}$, Jin Shang$^1$, Yujie Wang$^1$, and Jie Zhang$^{1,3,*}$}

\maketitle

\begin{affiliations}
	\item School of Physics and Astronomy, Shanghai Jiao Tong University, 800 Dong Chuan Road, Shanghai 200240, China
	\item Research Center for Advanced Science and Technology, University of Tokyo, 4-6-1 Komaba, Meguro-ku, Tokyo 153-8505, Japan
	\item Institute of Natural Sciences, Shanghai Jiao Tong University, Shanghai 200240, China
\end{affiliations}

\begin{abstract}
Experimental measurements of contact forces are limited to spheres and disks in three and two dimensions, making the evaluation of the shape effect and universality of force distributions and the comparison between experiments and theories extremely difficult. 
Here we present precise measurements of vector contact forces in photoelastic ellipses and disks subject to isotropic compression and pure shear. We find the local, instead of the global, stress ratio, control the width of the force distributions for forces larger than the mean, regardless of the particle shape and preparation protocols. 
By taking advantage of the anisotropic particle shape, we can determine the anisotropic growth of contacts in ellipses subject to isotropic compression, revealing the role of non-affine particle motions in homogenizing force distributions.
Our results uncover the role of local anisotropy in the statistical framework of granular materials and open a new regime of exploring the role of particle shape on the mechanical and dynamical properties of granular materials in depth.
\end{abstract}

\maketitle
\section*{Introduction}

Granular materials \cite{majmudar05nature} exhibit heterogeneous stress fields under external loading, as shared by emulsions \cite{brujic03,zhou06}, colloidal suspensions \cite{lin16colloids,yanagishima20}, nanoparticle packings \cite{lefever16}, and even biological cells \cite{tambe11}.
The force-chain particles of contact forces larger than the mean form the skeleton, playing key roles on the elasticity \cite{golderberg05}, the sound propagation \cite{jia99} and the plasticity \cite{wang20shearlocalization} of granular materials. 
The heterogeneity of stress distribution can be quantified using contact force distribution, which has become the focus of numerous experiments \cite{majmudar05nature,brujic03,zhou06,corwin05force} and simulations \cite{zhang05jamming,radjai96force,azema12force,boberski13,vanEerd07,hidalgo09,ohern01force,silbert02} over the last two decades. The asymptotic forms of normal contact force distribution $P(f)$ can be predicted using drastically different models, whose precise forms are still under intense debate. An exponential tail is predicted by the $q$-model \cite{liu95force}, the stress ensemble \cite{henkes2009filedTheory}, and the analogy of the spin glass \cite{edwards08force}, while a Gaussian tail is predicted by the force-network ensemble \cite{tighe10review}. Simulation results \cite{radjai96force,zhang05jamming,conzelmann20,azema12force,saitoh15,boberski13} are often inconsistent, and experimental results \cite{brujic03,zhou06,majmudar05nature} are limited in statistics.
 Nevertheless, the width of force distribution varies robustly with preparation protocols and particle shapes, even after the normalization of forces by the mean value. $P(f)$ decays faster, either by decreasing the global anisotropy \cite{majmudar05nature,vanEerd07}, or by isotropic compression \cite{hidalgo09,boberski13,conzelmann20}. Furthermore, $P(f)$ decays much slower for elongated particles in steady shear states \cite{azema12force}. Presently, a universal state variable that controls the width of force distribution is still missing.

Moreover, the contact-force measurements in experiments are still limited to spheres or disks, such as using two-dimensional (2D) photoelastic disks\cite{majmudar05nature,daniels17methods}, using the carbon paper \cite{mueth98carbon} or the photoelastic sheet \cite{corwin05force} on surfaces of 3D packings, and the bulk measurements of sphere packings \cite{brujic03,zhou06,brodu15,hurley16force,fischer21}. 
Since granular particles usually have anisotropic shapes, experimental force measurements on non-spherical or non-circular particles are needed to clarify the role of particle shape on the mechanical properties of granular materials. 

In this Article, we perform the first direct experimental measurement of the vector contact forces in ellipses and conduct a systematic study of the force distributions for ellipse packings and disk packings subject to isotropic compression (IC) and pure shear (PS). 
While the global stress anisotropy fails to capture the changes of normalized normal-force distributions $P(f_n^*)$, the width of $P(f_n^*)$ for $f_n^*>1$, i.e. forces larger than the mean, is proportional to the local stress anisotropy regardless of particle shapes and loading protocols. Furthermore, we reveal the mechanism of force homogenization by analysing the angular functions of averaged normal forces and contact numbers in ellipses subject to IC. Lastly, we clarify the relationship between the local stress anisotropy and the global stress anisotropy by analyzing the evolution of the spatial distributions of
particle principle stress orientations for ellipses and disks subject to PS.

\section*{Results}
\begin{figure}
	\centerline{\includegraphics[width = 8.6 cm]{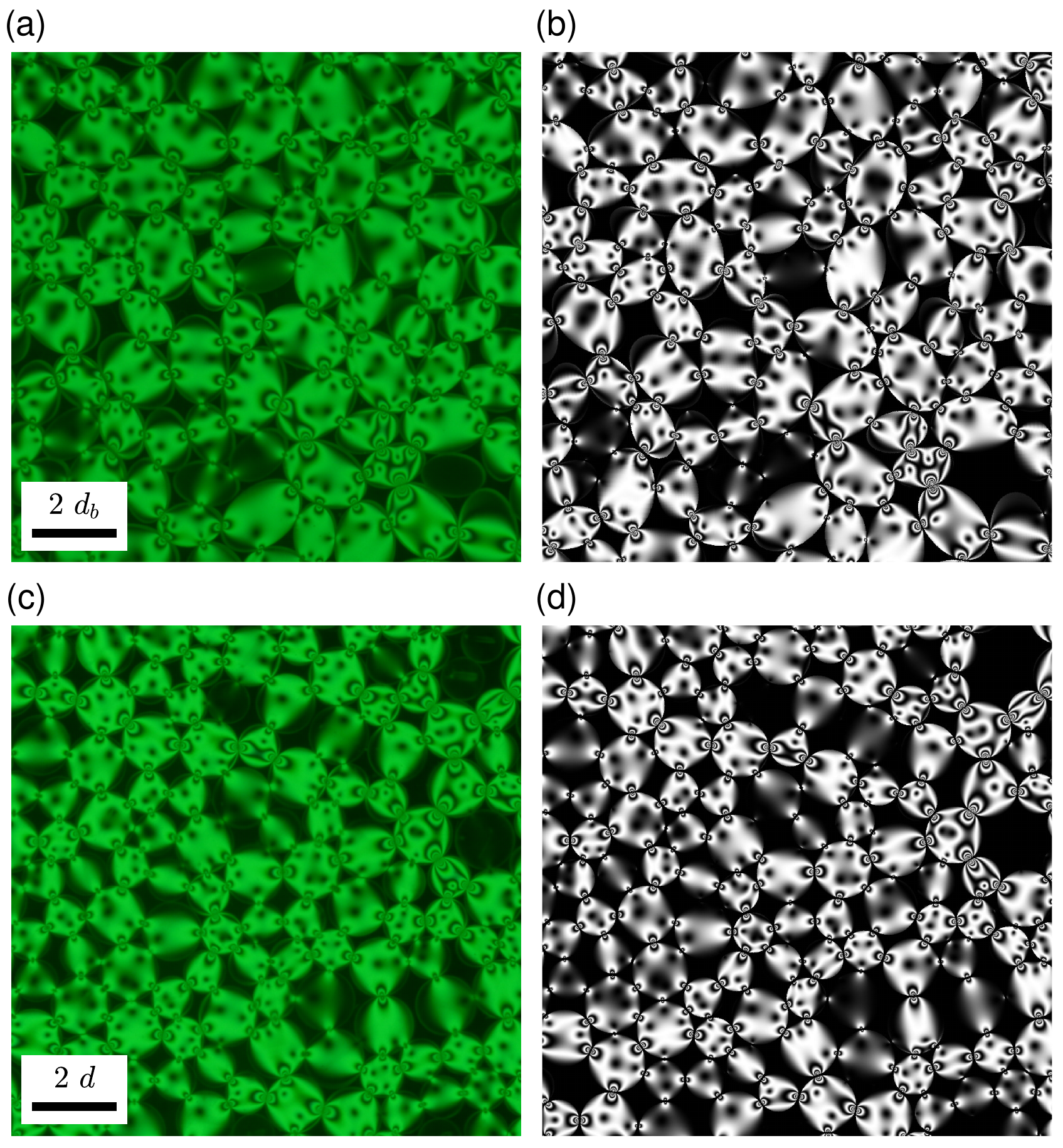}}
	\caption{\textbf{Comparison of the experimental stress images and the corresponding computed stress images using the measured contact forces.}
	Top row: a small portion of the bidisperse ellipse packing that consists of 1800 small and 900 large ellipses, whose aspect ratio is 1.5 with the minor axis $1.4~\rm{cm}$ for the large ellipse and $1.0~\rm{cm}$ for the small ellipse. On the scale bar, $d_b$ refers to the size of the minor axis of a small ellipse. Bottom row: a small portion of the bidisperse disk packing that consists of 2710 small disks and 1355 large disks, whose diameter is $1.4~\rm{cm}$ for a large disk and is $1.0~\rm{cm}$ for a small disk. On the scale bar, $d$ refers to the small-disk diameter. (a) and (c) The experimental stress images taken with green polarized light. (b) and (d) The computed stress images using the measured vector contact forces.}
	\label{fig:figure1}
\end{figure}

\paragraph*{Experimental details.}
We use a biaxial apparatus to apply IC or PS on 2D granular systems \cite{wang20shearlocalization}. The apparatus mainly consists of a rectangular frame filled with a layer of particles on a glass plate. The two pairs of walls can move symmetrically to apply IC or can move with one pair compression and another pair expansion to apply PS with a motion precision of 0.1 mm while keeping the center of mass fixed. Below the glass plate, eight mini vibrators are attached to eliminate the friction between the particle layer and the glass plate. 

For both photoelastic ellipses and disks, we record the normal image of particle configuration to detect the positions and orientations of particles. We also record the stress image to measure contact forces, as shown in Fig.~\ref{fig:figure1}(a), (c).
Vector contact forces are analyzed from the stress image using a force-inverse algorithm, which generates a computed stress image based on an initial guess of the contact forces, and then iterate to minimize the difference between the experimental and computed stress image.
Developed from the early version on disks \cite{majmudar05nature}, our algorithm can measure vector contact forces of ellipses with possible extension to other 2D particle shapes, see details in the Supplemental Information.
The relative error of contact force measurement is less than 5\% for the typical force magnitude, and the accuracy of measured contact forces can be visually checked by plotting the computed images for comparison as shown in Fig.~\ref{fig:figure1}.

\begin{figure}
	\centerline{\includegraphics[width = 8.6 cm]{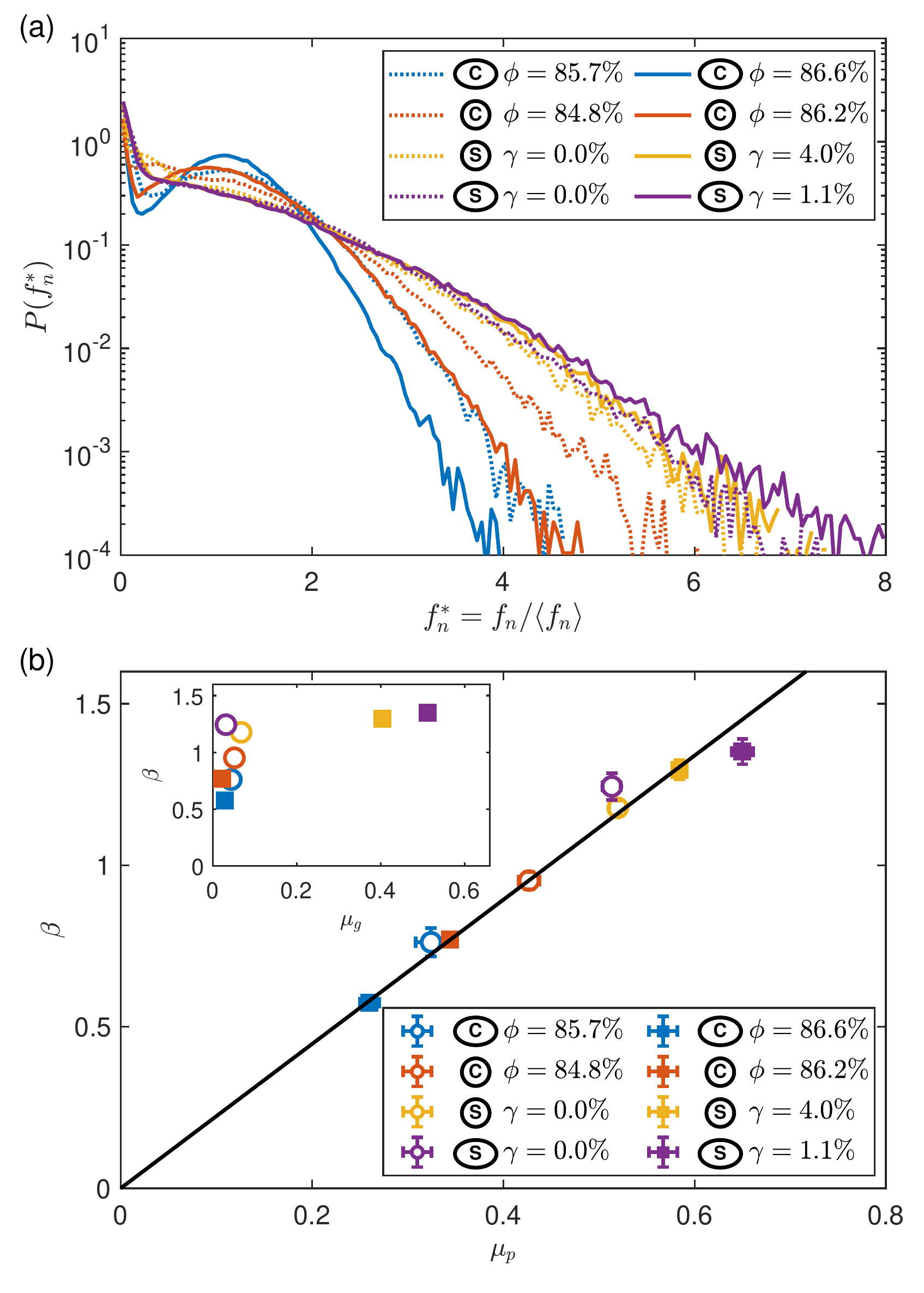}}
	\caption{\textbf{Probability density functions of the normal forces and their widths $\beta$ versus the average particle stress ratios $\mu_p$.} (a) PDFs of the normalized normal forces $f_n^*=f_n/\langle f_n \rangle$, where $\langle f_n \rangle$ refers to the mean normal force. In the legend, the circle/ellipse markers denote the experiments using the circular/elliptical photoelastic particles, and the characters `\textbf{C}' and `\textbf{S}' within the markers denote the different experimental protocols of IC, i.e. `\textbf{C}', and PS, i.e. `\textbf{S}'. The symbol $\phi$ denotes the packing fraction and $\gamma$ denotes the shear strain. In the shear experiments, $\phi = 85.0\%$ for disk packings and $\phi = 85.2\%$ for ellipse packings. (b) The widths of $P(f_n^*)$ versus the particle stress ratios $\mu_p$. The black straight line represents the linear fitting of the data without an intercept term. The error bars denote the standard deviations of at least 30 independent runs. Inset: The widths $\beta$ versus the global stress ratios $\mu_g$.}
	\label{fig:figure2}
\end{figure}

\paragraph*{Normalized normal-force distributions $P(f_n^*)$.}
To study the force distributions systematically, we apply two types of loading of IC and PS on the ellipse and disk packings.
For IC, we prepare the stress-free initial states near random close packings, whose packing fraction $\phi$ is slightly below the jamming point of frictionless particles. The initial $\phi$ is 83.4\% for disk packings and is 84.7\% for ellipse packings. 
Starting from the initial stress-free states, the four walls move inwards symmetrically in steps. At the same time, vibrators are turned on. 
For PS, first, we apply cyclic shear on a jammed packing to drive it into the steady state of cyclic shear. The $\phi$ is 85.0\% for disk packings and is 85.2\% for ellipse packings, respectively. The amplitude of cyclic shear strain is 2.8\% for disk packings and is 1.1\% for ellipse packings, respectively. Then we apply PS quasi-statically in steps by compressing along the $x$ axis and expanding along the $y$ axis while keeping the area fixed. The zero shear strain $\gamma=0$ reference states are chosen as the global stress ratios $\mu_g$ vanish.

For two particle types, two protocols, and two typical levels of compression or shear, we analyse the normal-force distributions of eight jammed states, as shown in Fig.~\ref{fig:figure2}(a). The typical stress images of the eight jammed states are shown in Supplementary Fig. 9 .
The magnitude of the normal force $f_n$ is normalized by its mean value $\langle f_n \rangle$, and $P(f_n^*)$ are ensemble averaged over 30 independent runs.

Subject to IC, $P(f_n^*)$ shows a peak around $f_n^*=1$ and decays rapidly as $\phi$ increases, regardless of the particle shape. Compared to disks, ellipses have larger $\phi$ and can reach more homogeneous states in view of force distributions.

For the anisotropic jammed states, $P(f_n^*)$ of both ellipses and disks decay much slower than those of IC. $P(f_n^*)$ of ellipses decays faster than that of disks under IC, whereas $P(f_n^*)$ of ellipses decays slower than that of disks under PS. Moreover, $P(f_n^*)$ loses its peak around the mean value, suggesting the presence of a peak in $P(f_n^*)$ may not be a robust signature of jammed state \cite{ohern01force,corwin05force}.

Both the particle shapes and protocols show significant influence on the shape of force distributions, even after the normalization of forces by the mean value. The previous experiments \cite{majmudar05nature} and simulations \cite{vanEerd07} report that when the global stress anisotropy decreases, $P(f_n^*)$ changes from an exponential tail to a Gaussian tail. Decompressing an isotropic jammed state, $P(f_n^*)$ can decay even faster than Gaussian \cite{vanEerd07}. In experiments, we find that it is difficult to determine the explicit asymptotic form of $P(f_n^*)$ since the different models provide equally good fitting for $f_n^*>1$, possibly due to the finite range of statistics in force measurements, as shown in Supplementary Fig.10 . 

Nevertheless, a robust feature is the width of $P(f_n^*)$ for $f_n^*>1$ do change with the protocols and particle shapes, as
quantified using 
\begin{equation}
    \beta\equiv\sqrt{\frac{1}{N_c}\sum_{i>j}(f_{n,ij}^*-1)^2}, \quad f_{n,ij}^* > 1.
\end{equation}
Here $f_{n,ij}^*$ is the normalized normal force between particle $i$ and particle $j$, $N_c$ is the number of contacts whose forces are larger than the mean value.

The previous experiments \cite{majmudar05nature} and simulations \cite{vanEerd07} report that, increasing the global stress anisotropy can broaden $P(f_n^*)$, which, however, clearly fails to account for the differences of $P(f_n^*)$ in the isotropic jammed states of different compression levels and also of different particle shapes. 

\paragraph*{Average particle stress ratio $\mu_p$.}To reveal the elusive control parameter for the width $\beta$, we first define the particle virial stress tensor, $\boldsymbol{\sigma}_i = \frac{1}{S_i}\sum_j\Vec{f}_{ij}\otimes\Vec{r}_{ij}$.
Here $S_i$ is the Voronoi area of particle $i$, $\Vec{f}_{ij}$ is the contact force vector between the particles $i$ and $j$, $\Vec{r}_{ij}$ is the vector from the center of particle $i$ to the contact point of the particles $i$ and $j$. The  '$\otimes$' represents the dyadic product. 
The particle stress ratio is given as, 
\begin{equation}
  \mu_{i} = \frac{\tau_i}{p_i}=\left | \frac{\sigma_{1,i}-\sigma_{2,i}}{\sigma_{1,i}+\sigma_{2,i}} \right |,
\end{equation}
where $\sigma_{1,i}$ and $\sigma_{2,i}$ are the two eigenvalues of $\boldsymbol{\sigma}_i$.
Then, we define the average particle stress ratio $\mu_p$ as,
\begin{equation}
\mu_p = \frac{1}{N}\sum_{i=1}^N\mu_i,
\end{equation}
where $i$ runs over particles in the force network, whose pressure $p_i$ is larger than the mean value $\langle p \rangle$. Thus, the $\mu_p$ quantifies the local stress anisotropy of the packing.

We find that it is the particle stress ratio $\mu_p$, instead of the global stress ratio $\mu_g$, that controls the width of the normal-force distributions $P(f_n^*)$. As shown in Fig.~\ref{fig:figure2}(b), the widths $\beta$ of $P(f_n^*)$ are just proportional to the particle stress ratios $\mu_p$. The black straight line is the linear fitting of the data without an intercept term. We plot the widths $\beta$ versus the global stress ratios $\mu_g$ in the inset in Fig.~\ref{fig:figure2}(b), which shows that $\mu_g$ can not determine the width $\beta$.
Influenced by the particle shapes and the preparation protocols, the systematic change of the tangential force distributions is much more complex, as shown in Supplementary Fig. 11 , though the general feature of the exponential tails is consistent with the early observation \cite{majmudar05nature}.

\begin{figure}
	\centerline{\includegraphics[width = 8.6 cm]{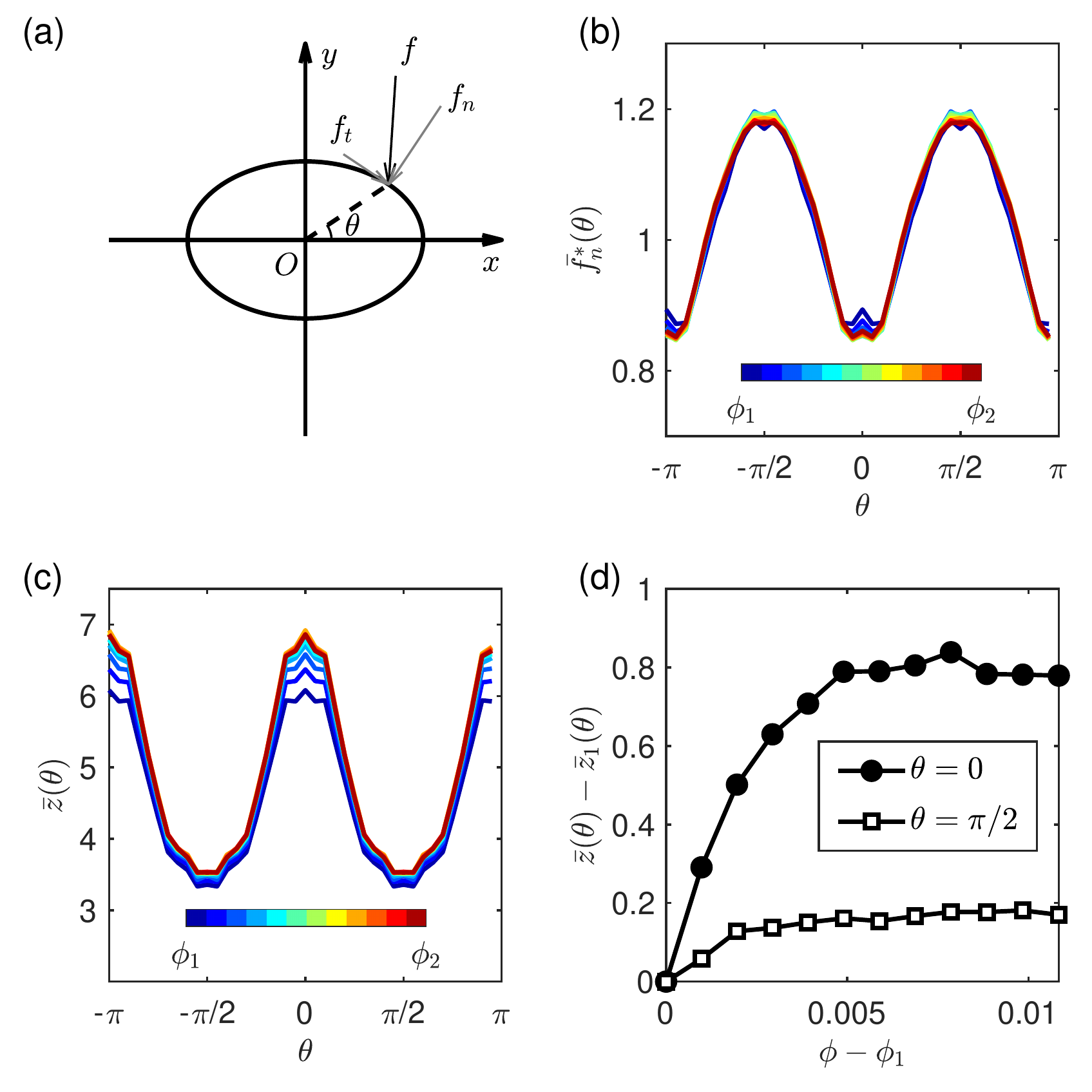}}
	\caption{\textbf{Evolution of the angular functions of the averaged normal forces $\bar{f_n^*}(\theta)$ and the averaged contact numbers $\bar{z}(\theta)$ for ellipse packings subject to isotropic compression.} (a) Illustration of the contact angle $\theta$ and the vector contact force decomposition for an ellipse, $f_n$ is the normal force and $f_t$ is the tangential force. Angular functions of (b) the averaged normalized normal forces $\bar{f_n^*}(\theta)$ and (c) the averaged contact numbers $\bar{z}(\theta)$ of different compression levels, where $\phi_1=85.7\%$ and $\phi_2=86.6\%$. (d) Evolution of contact numbers $\bar{z}(\theta)$ for $\theta=0$ and $\theta=\pi/2$.}
	\label{fig:figure3}
\end{figure}

Next, we will focus on the evolution of the force network subject to IC and PS and clarify the changes of $P(f_n^*)$ in detail.

\paragraph*{Evolution of force network subject to IC.}
Under IC, despite the mean value of $f_n$ increases, the width of the normalized force distributions $P(f_n^*)$ decreases for both disk packings and ellipse packings. This corresponds to a more homogeneous force network. 
The homogenization of the force network indicates the non-affine motions of particles during the compression. In experiment, it is difficult to detect the reliable particle non-affine displacements during IC due to the finite resolution. Owing to the anisotropy shape of ellipses, we can use the birth of contacts to characterize the non-affine motions of particles, which leads to a more uniform force network.

As shown in Fig.~\ref{fig:figure3}(b), the angle dependence of the averaged normal forces show tiny changes subject to IC. The ratio between the magnitude of normal forces near the minor axis $\theta=\pi/2$ and the major axis $\theta=0$ is about 1.4, close to the aspect ratio $\eta=1.5$ of the ellipse. Fig.~\ref{fig:figure3}(c) shows that the averaged contact number near the major axis is much larger than the corresponding value near the minor axis. In contrast to the small changes of the averaged normal forces shown in Fig.~\ref{fig:figure3}(b), the contact numbers increase significantly during compression with a clear dependence on the contact angle $\theta$, as shown in Fig.~\ref{fig:figure3}(c), (d).   
Contacts prefer to develop near the major axis of an ellipse. This anisotropic increase of contact numbers reveals the non-affine motions of particles subject to IC. On the other hand, the averaged normal force near the major axis is smaller than the mean value. Therefore, the contact prosperity near the major axis will increase the weight of small forces, and consequently decrease the width $\beta$ of force distributions.
Furthermore, the orientation $\theta_p$ and the principle-stress orientation $\theta_s$ of particles show strong correlations under IC, which become much less pronounced under shear, as shown in Supplementary Fig. 12 .

\begin{figure}
	\centerline{\includegraphics[width = 8.6 cm]{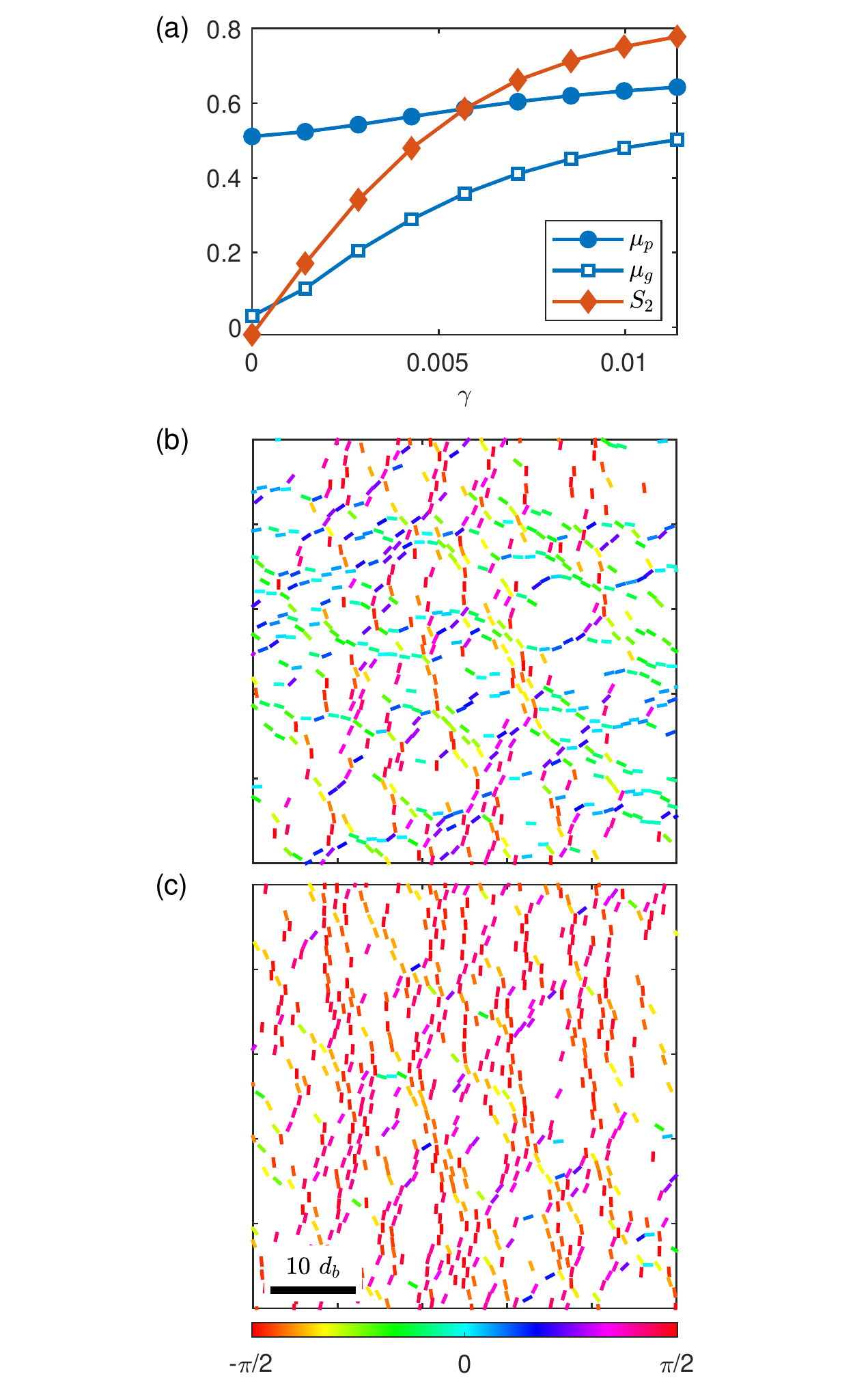}}
	\caption{\textbf{Evolution of stress ratios and the particle principle stress orientations for ellipse packings subject to pure shear.} (a) The global stress ratios $\mu_g$, the particle stress ratios $\mu_p$ and the nematic order of the particle principle stress orientations $S_2$ as a function of the shear strain $\gamma$. The spatial distributions of the particle principle stress orientations in (b), where $\gamma=0$, and in (c), where $\gamma=0.011$. The painted bars denote the particle principle stress orientations. Only particles with pressure larger than the mean value are shown.}
	\label{fig:figure4}
\end{figure}

\paragraph*{Evolution of force network subject to PS.}
Under PS, the global stress ratio $\mu_g$ undergoes a significant increase from almost zero to $\mu_g \approx 0.4$ for disk packings, and from almost zero to $\mu_g \approx 0.5$ for ellipse packings. However, correspondingly, $P(f_n^*)$ at $\gamma=0.0\%$ and $\gamma=1.1\%$ for ellipse packings show small changes, and the similar observation holds for disk packings, which is nonetheless consistent with the small changes of the local stress ratios $\mu_p$ shown in Fig.~\ref{fig:figure4}(a).

To gain an in-depth understanding of the differences and interrelationships between the global and local stress ratios, we plot the evolution of force network under shear, as shown in Fig.~\ref{fig:figure4}(b), (c). The force-chain particles are drawn as painted bars, representing the orientation of the principle stress $\theta_s$ of particle. 
In an isotropic state, $\theta_s$ distribute randomly, as shown in Fig.~\ref{fig:figure4}(b), and show weak correlations with the particle orientations $\theta_p$, as shown in Supplementary Fig. 12(c). 
Therefore, the local stress anisotropy cancels with each other during the global average, leading to a vanishing global stress ratio $\mu_g$. Subject to PS, the principle stress of particle trends to align with the compression direction, as shown in Fig.~\ref{fig:figure4}(c), thus totally decorrelated with the particle orientations $\theta_p$, as shown in Supplementary Fig. 12(d) . 
The alignment can be quantified using the nematic order $S_2 = \langle2\cos^2(\theta_s-\frac{\pi}{2})-1\rangle$ of the particle principle stress orientations, as plotted in Fig.~\ref{fig:figure4}(a). Consequently, the global stress ratio $\mu_g$ can make a large change by incorporating the alignment of the particle principle stress with the little increase of $\mu_p$. 
The similar behaviors are also seen in disks, as shown in Supplementary Fig. 13.

\section*{Discussion and conclusion}
It is well-known that loading protocols can change the shape of force distributions, even normalized by the mean force. 
Our results point out that the average particle stress ratio $\mu_p$ is the actual control parameter, which varies with loading protocols and material properties of particles, such as shape and friction. The width $\beta$ of $P(f_n^*)$ 
can be related to $\langle (f_n^*)^2 \rangle$ with a reasonable approximation, which is regarded as an additional constraint in deriving the stress distribution subject to the entropy maximization in force-network ensemble \cite{tighe10FNE,bilign19protocol} given that $\langle (f_n^*)^2\rangle \sim \langle p^2 \rangle$. Our results provide a clear physical meaning that the magnitude of the constraint $\langle (f_n^*)^2 \rangle$ is related to the local stress anisotropy $\mu_p$. Moreover, it is surprising that $\mu_p \propto \beta$ is independent of particle shape, since in ellipses $f_n^*$ is not only related to the force balance but also is related to the torque balance, which would imply one more constraint in the force tiling.

In summary, we perform the first direct experimental measurement of vector contact forces in ellipses, permitting a comprehensive analysis and the comparison of vector contact forces in both ellipses and disks. By studying $P(f_n^*)$, we find the particle stress ratios $\mu_p$ control the width of $P(f_n^*)$ for $f_n^* > 1$, regardless of the particle shapes and protocols.
Looking into future, we are now at a good position to address many outstanding physics and engineering problems, which include the jamming problem of ellipses and the possible extension of force measurement to the other particle shapes in two-dimensions and to build connection between microscopic mechanics of particle properties to the macroscopic responses of system. Our present work clearly opens a new regime of exploring the role of particle shape on the mechanical and dynamical properties of granular materials in depth.

\begin{addendum}
	\item This work is supported by the NSFC (No.11974238 and No.11774221 ). This work is also supported by the Innovation Program of Shanghai Municipal Education Commission under No 2021-01-07-00-02-E00138. We also acknowledge the support from the Student Innovation Center of Shanghai Jiao Tong University.
	
	\item[Competing Interests] The authors declare that they have no competing financial interests.
	
	\item[Author Contributions] J.Z. conceived and supervised the project. Y.Q.W. and J.Z. designed the experiment. Y.Q.W. and J.S. cut the photoelastic ellipses and performed the experiment. Y.Q.W. developed the force-inverse algorithm of ellipses and analysed the data. Y.Q.W., J.S., Y.J.W. and J.Z. participated in the discussions. Y.Q.W. and J.Z. wrote the paper.
	\item[Correspondence and requests for materials] should be addressed to J.Z. (email: jiezhang2012@sjtu.edu.cn).
\end{addendum}

\end{document}


\title{Supplemental Information: Contact force measurements and local anisotropy in ellipses and disks}

\author{Yinqiao Wang}
    \affiliation{School of Physics and Astronomy, Shanghai Jiao Tong University, 800 Dong Chuan Road, 200240 Shanghai, China.}
    \affiliation{Research Center for Advanced Science and Technology, University of Tokyo, 4-6-1 Komaba, Meguro-ku, Tokyo 153-8505, Japan.}
\author{Jin Shang}
    \affiliation{School of Physics and Astronomy, Shanghai Jiao Tong University, 800 Dong Chuan Road, 200240 Shanghai, China.}
    
\author{Yujie Wang}
    \affiliation{School of Physics and Astronomy, Shanghai Jiao Tong University, 800 Dong Chuan Road, 200240 Shanghai, China.}
    
\author{Jie Zhang}
    \email[Email address: ]{jiezhang2012@sjtu.edu.cn}
    \affiliation{School of Physics and Astronomy, Shanghai Jiao Tong University, 800 Dong Chuan Road, 200240 Shanghai, China.}
    \affiliation{Institute of Natural Sciences, Shanghai Jiao Tong University, 200240 Shanghai, China.}

\maketitle

\section{Photoelastic methods}
\subsection{Setup of the contact-force measurements and the general philosophy of the force-inverse algorithm}

The contact-force measurements using the photoelastic techniques \cite{MajmudarThesis,Daniels2017,WangThesis} are based on the stress-induced birefringence, resulting in a stress image of a bright and dark fringe pattern within the photoelastic sample material as visualized through a pair of polarizers. 

The description of the related phenomenon can be traced back to as early as a paper by Brewster in 1815 \cite{Bre1815}. This phenomenon was later further elaborated by Coker and Filon in 1930 \cite{Coker1931}. The first application of photoelasticity as a measurement technique in the field of granular materials began in 1950 with the work of Wakabayashi \cite{waka1950}, which was soon followed by the independent work of Drescher, Oda, and de Jong and others for experimental measurements of granular materials \cite{Drescher1972,Drescher1976,Oda,Oda1974}.

A schematic diagram of the polariscope is shown in Supplementary Fig. \ref{fig:setup}(a) and an example of a stress image of a disk subject to two vector contact forces is shown in Supplementary Fig. \ref{fig:setup}(b).

\begin{figure}[htp!]
  \centering
  \includegraphics[width=15cm]{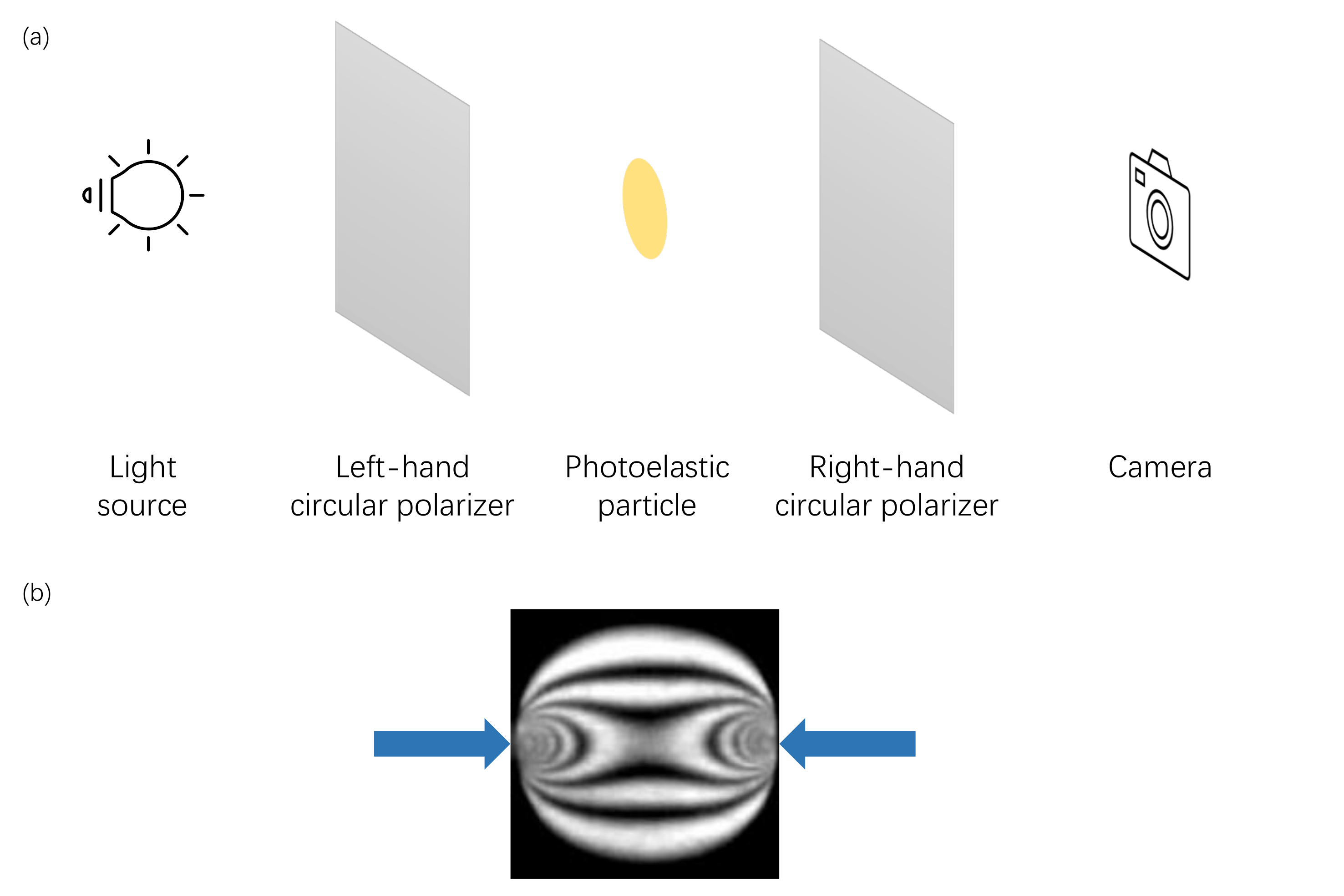} 
  \caption {(a) Schematic diagram of the circular dark-field transmission polariscope. From the left to the right, a light source, a left-hand circular polarizer, a photoelastic particle, a right-hand circular polarizer and a camera. The quarter-wave plates of the two circular polarizers face the particle. (b) The stress image of a disk subject to two contact forces.}
 \label{fig:setup}
\end{figure}

The intensity $I^e$ of the experimental stress image in the photoelastic sample material, whose shape can be disk, ellipse, or any other shapes, is given by,
\begin{equation}
    I^e(x,y)=I_0\sin^2\left ( \frac{\pi h C}{\lambda}\left(\sigma_1-\sigma_2\right) \right),
    \label{eq:raw_stress_intensity}
\end{equation}
where $h$ is the sample thickness, $\lambda$ is the wavelength of light, $\sigma_1$ and $\sigma_2$ are the two eigenvalues of the two-dimensional stress tensor at any given point $(x,y)$ within the sample, $C$ is the stress-optic coefficient, and $I_0$ is a constant.
In the above equation, the parameters $h$, $\lambda$, $C$, and $I_0$ are either readily available from the material properties or can be determined by calibration.

We want to emphasize that the problem is essentially two dimensional(2d) such that a disk or an ellipse can be treated as a 2d object with $I^e$ and $\sigma_1$ and $\sigma_2$ being defined within the $(x,y)$ plane of the disk or the ellipse. Therefore, all we discuss in the following will be essentially 2d.

We also want to emphasize that $\sigma_1$ and $\sigma_2$ in eq.~(\ref{eq:raw_stress_intensity}) are the implicit functions of the contact force vectors $\{f_{x,i},f_{y,i}\}$ at every contact point $i$ on the particle, whose shape can be disk, ellipse, or any other shapes.
Therefore, the general philosophy of the force-inverse algorithm is to do an inverse engineering to go from the function $I^e(x,y)$ of the experimental stress image on the $l.h.s.$ of eq.~(\ref{eq:raw_stress_intensity}) to find the contact force vectors $\{f_{x,i},f_{y,i}\}$ within the implicit functions $\sigma_1$ and $\sigma_2$ on the $r.h.s$ of eq.~(\ref{eq:raw_stress_intensity}).

\begin{figure}
  \centering
  \includegraphics[width=15cm]{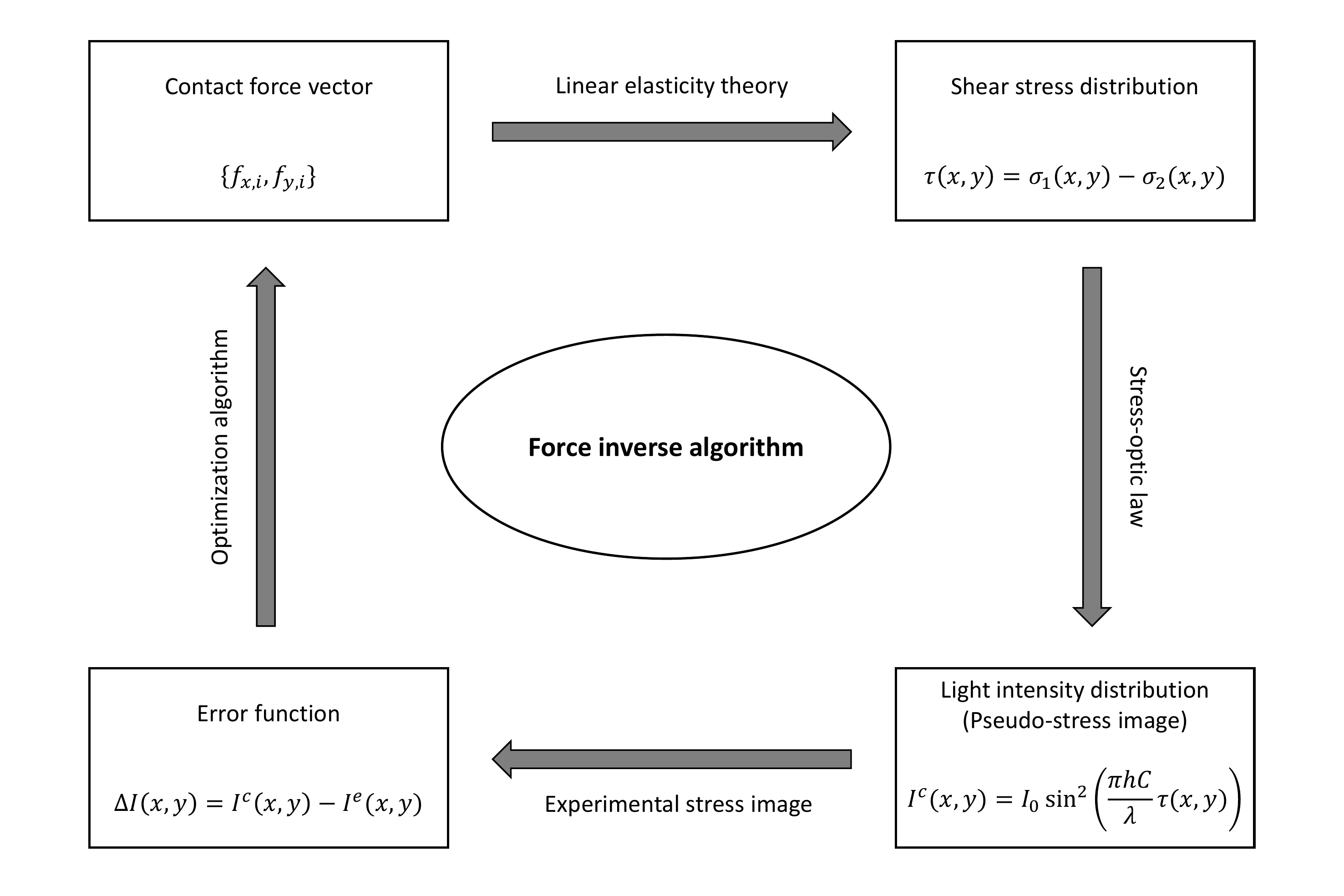} 
  \caption {Illustration of the force-inverse algorithm.}
 \label{fig:ForceInverse-1}
\end{figure}

The key elements of our force-inverse algorithm are shown in Supplementary Fig.\ref{fig:ForceInverse-1}. This algorithm generates a computed stress image based on an initial guess of the contact forces, and then iterates to minimize the difference between the experimental stress image and the corresponding computed stress image, as shown in Supplementary Fig. \ref{fig:ForceInverse-1}. The framework of the algorithm was first proposed by Majmudar and Behringer in their pioneering work of the photoelastic force measurement in disks in 2005 \cite{MajmudarThesis}. We will show below that this framework works for the photoelastic force measurement in both disks and ellipses.

\begin{figure}
  \centering
  \includegraphics[width=12cm]{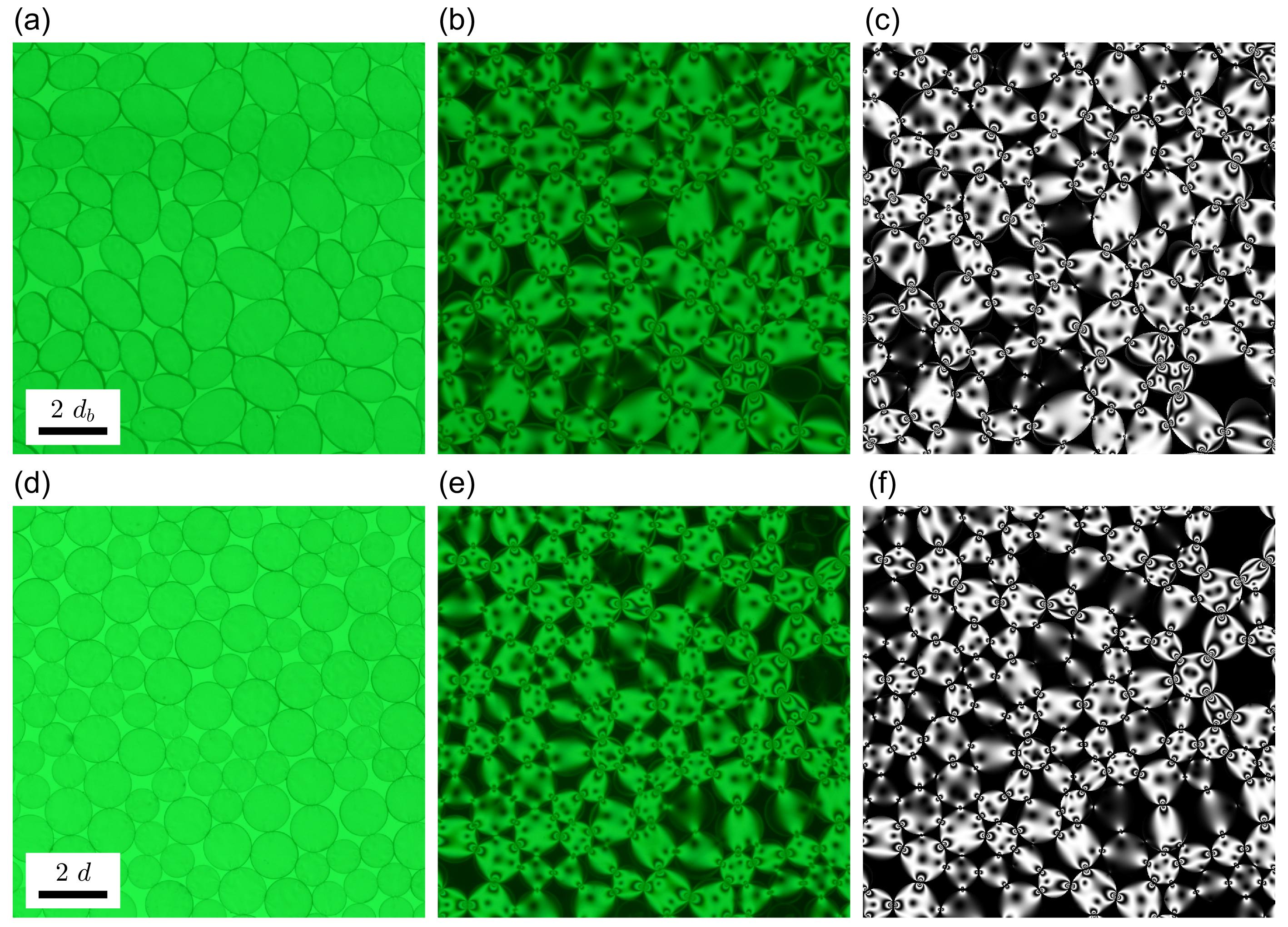} 
  \caption {(a) and (d) The normal images. (b) and (e) The experimental stress images. (c) and (f) The computed stress image.}
 \label{fig:nscimg}
\end{figure}

In the experiments, we capture two types of images, the normal images without polarizers and the stress images with polarizers, as shown in Supplementary Fig. \ref{fig:nscimg}. We first detect the particle positions and the contact points using the normal images, and then with the stress images we can solve the vector contact forces of disks using our force-inverse algorithm. 

\subsection{Contact-force measurements of disks}
To compute the stress image within a photoelastic particle with the contact force vectors, we need the stress solutions of the linear elastic theory. For a disk, the solutions are analytical \cite{MajmudarThesis, Daniels2017}. Firstly, with the contact force vectors $\{f_{x,i},f_{y,i}\}$ at every contact point $i$ on the disk, we can solve the stress distribution within the disk by applying the analytical solutions of the 2d linear elastic theory of a disk, which are,
\begin{equation}
\begin{split}
    \sigma_{xx}(x,y)&=\sum_{i=1}^z \frac{-2}{\pi h}\frac{(x-x_i)^2\left[(x-x_i)f_{x,i}+(y-y_i)f_{y,i}\right]}
    {\left[(x-x_i)^2+(y-y_i)^2\right]^2} \\
    \sigma_{yy}(x,y)&=\sum_{i=1}^z \frac{-2}{\pi h}\frac{(y-y_i)^2\left[(x-x_i)f_{x,i}+(y-y_i)f_{y,i}\right]}
    {\left[(x-x_i)^2+(y-y_i)^2\right]^2} \\
    \sigma_{xy}(x,y)&=\sum_{i=1}^z \frac{-2}{\pi h}\frac{(x-x_i)(y-y_i)\left[(x-x_i)f_{x,i}+(y-y_i)f_{y,i}\right]}
    {\left[(x-x_i)^2+(y-y_i)^2\right]^2}. 
\end{split}
    \label{eq:force_stress}
\end{equation}
Here $(x_i,y_i)$ are the coordinates of the contact point $i$, $(f_{x,i},f_{y,i})$ are the corresponding force vector components, and $(x,y)$ refer to the coordinates of any point of within the disk, as illustrated in Supplementary Fig.~\ref{fig:ForceInverse-2}. Note that $z$ is the contact number, which shall not be confused with the usual three-dimensional component. $\sigma_{\alpha\beta}$ represent the components of the 2d stress tensor.
From eq.~(\ref{eq:force_stress}), we can obtain the maximum shear stress of any point,
\begin{equation}
    \sigma_1-\sigma_2 = \sqrt{(\sigma_{xx}-\sigma_{yy})^2+4\sigma_{xy}^2}.
    \label{eq:shear_stress}
\end{equation}

\begin{figure}
  \centering
  \includegraphics[width=15cm]{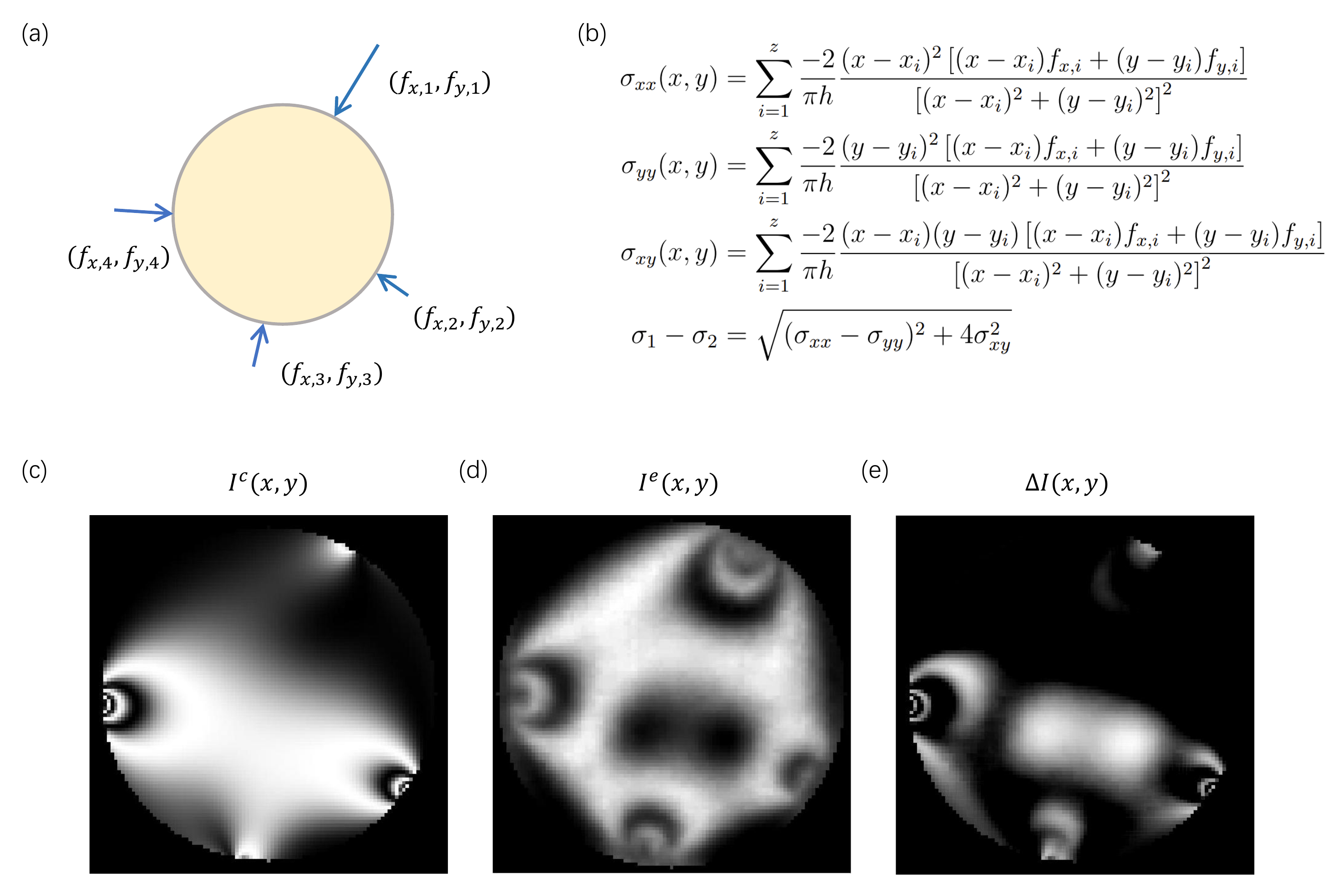} 
  \caption {(a) A diagram of a disk subject to four vector contact forces. (b) The analytical stress solutions of a disk. (c) The computed stress image following an initial guess of contact forces. (d) The experimental stress image. (e) The difference of the computed stress image and the experimental stress image.}
 \label{fig:ForceInverse-2}
\end{figure}

Second, the stress distribution of the disk results in the computed (pseudo) stress image via the stress-optic law. The intensity of the computed stress image $I^c$ is given by,
\begin{equation}
    I^c=I_0\sin^2\left ( \frac{\pi h C}{\lambda}\left(\sigma_1-\sigma_2\right) \right),
    \label{eq:stress_intensity}
\end{equation}
which is essentially the substitution of eq.~(\ref{eq:shear_stress}) into the $r.h.s$ of eq.~(\ref{eq:raw_stress_intensity}).    

\begin{figure}
  \centering
  \includegraphics[width=15cm]{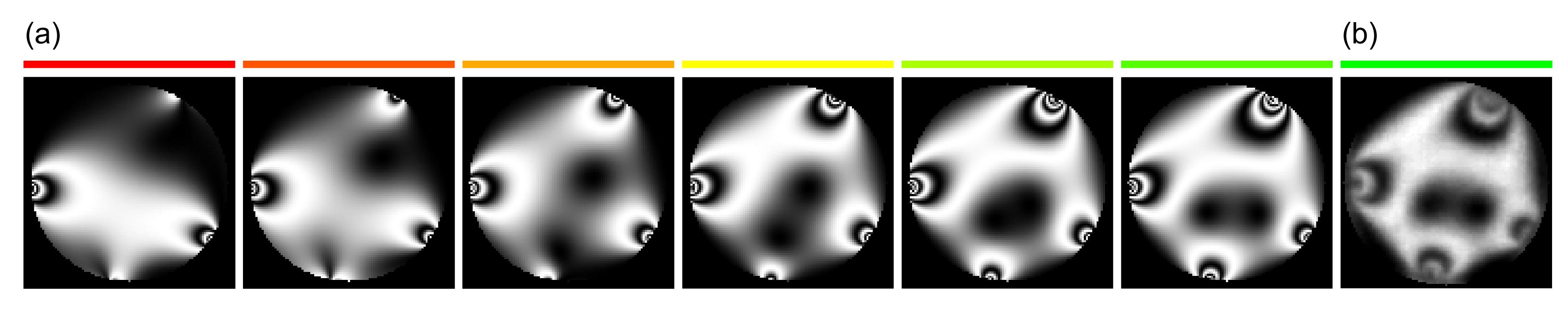} 
  \caption {An example of the converging process of the force-inverse algorithm. (a) A series of reconstructed stress images from a preliminary estimation of contact force vectors to the most optimized contact force vectors. (b) The experimental stress image captured by the camera.}
 \label{fig:simgSeries}
\end{figure}

Compared to the experimental stress image $I^s$, we can define the error function as the difference of the computed stress image and the experimental stress image, as illustrated in Supplementary Fig.\ref{fig:ForceInverse-2}. Finally, the optimization algorithm will minimize the error function according to adjusting the input contact-force vectors $\{f_{x,i},f_{y,i}\}$. Supplementary Figure \ref{fig:simgSeries} shows the process of the computed stress image converging to the experimental stress image. Once the error function vanishes, the input contact force vectors give the calculated results.

\subsection{Contact force measurements of ellipses}

For non-circular particles, the key point is that the analytical solutions of the stress distribution with the known contact force vectors are often unavailable. As a result, no much progress has been made in the vector contact force measurement for non-circular particles since the pioneering work of Majmudar and Behringer in 2005 \cite{MajmudarThesis}. 

\begin{figure}
  \centering
  \includegraphics[width=15cm]{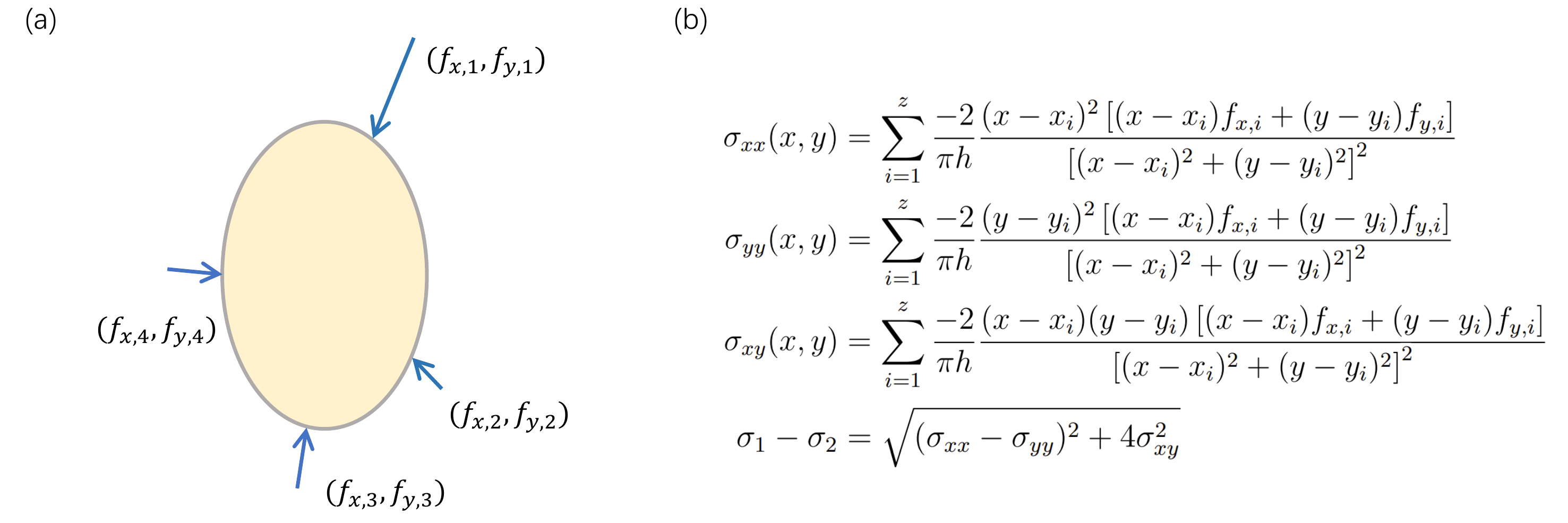} 
  \caption {(a) A diagram of an ellipse subject to four vector contact forces. (b) The approximate analytical solutions of an ellipse using the analytical stress solutions of a disk.}
 \label{fig:ForceInverse-3}
\end{figure}

To overcome this difficulty, we propose two methods to obtain the stress distributions of ellipses with the known contact force vectors: the approximate method using the analytical solutions of disks, see eq. (\ref{eq:force_stress}) and Supplementary Fig.\ref{fig:ForceInverse-3}, and the numerical solution using the Finite-Element-Method \cite{WangThesis}. The latter method can be implemented using the Partial Differential Equation Toolbox in MATLAB \cite{WangThesis}. 

\begin{figure}
  \centering
  \includegraphics[width=14.8cm]{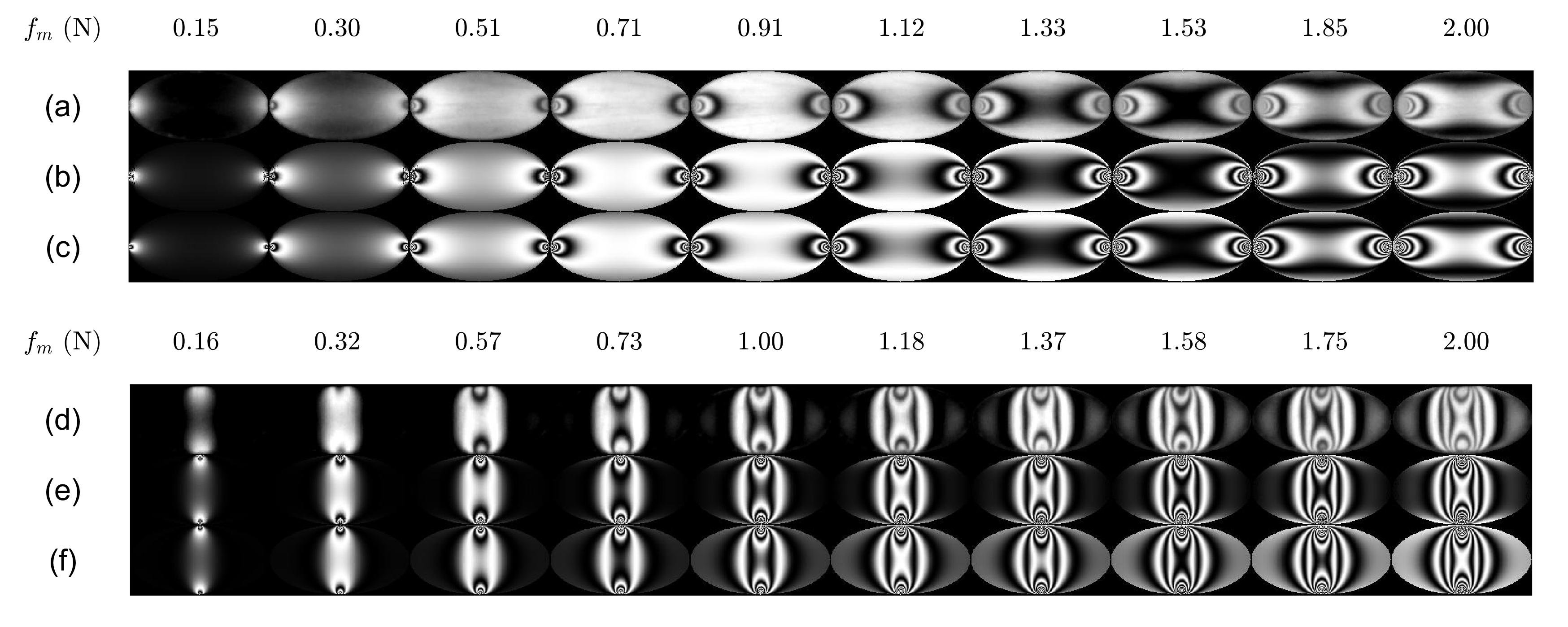} 
  \caption {Comparison of the experimental stress images captured by the camera and the corresponding computed stress images reconstructed using the two force-inverse algorithms for the ellipse ($r_a=$ 1.0 cm, $r_b=$ 0.5 cm). Major axis loading: (a) experimental stress images, (b) computed stress images using the FEM method, (c) computed stress images using the approximate method. Minor axis loading: (d) experimental stress images, (e) computed stress images using the FEM method, (f) computed stress images using the approximate method. The force magnitudes measured by a force gauge $f_m$ in the units of Newtons are labeled above individual images.}
 \label{fig:ellipse_force_image}
\end{figure}

\begin{figure}
  \centering
  \includegraphics[width=8cm]{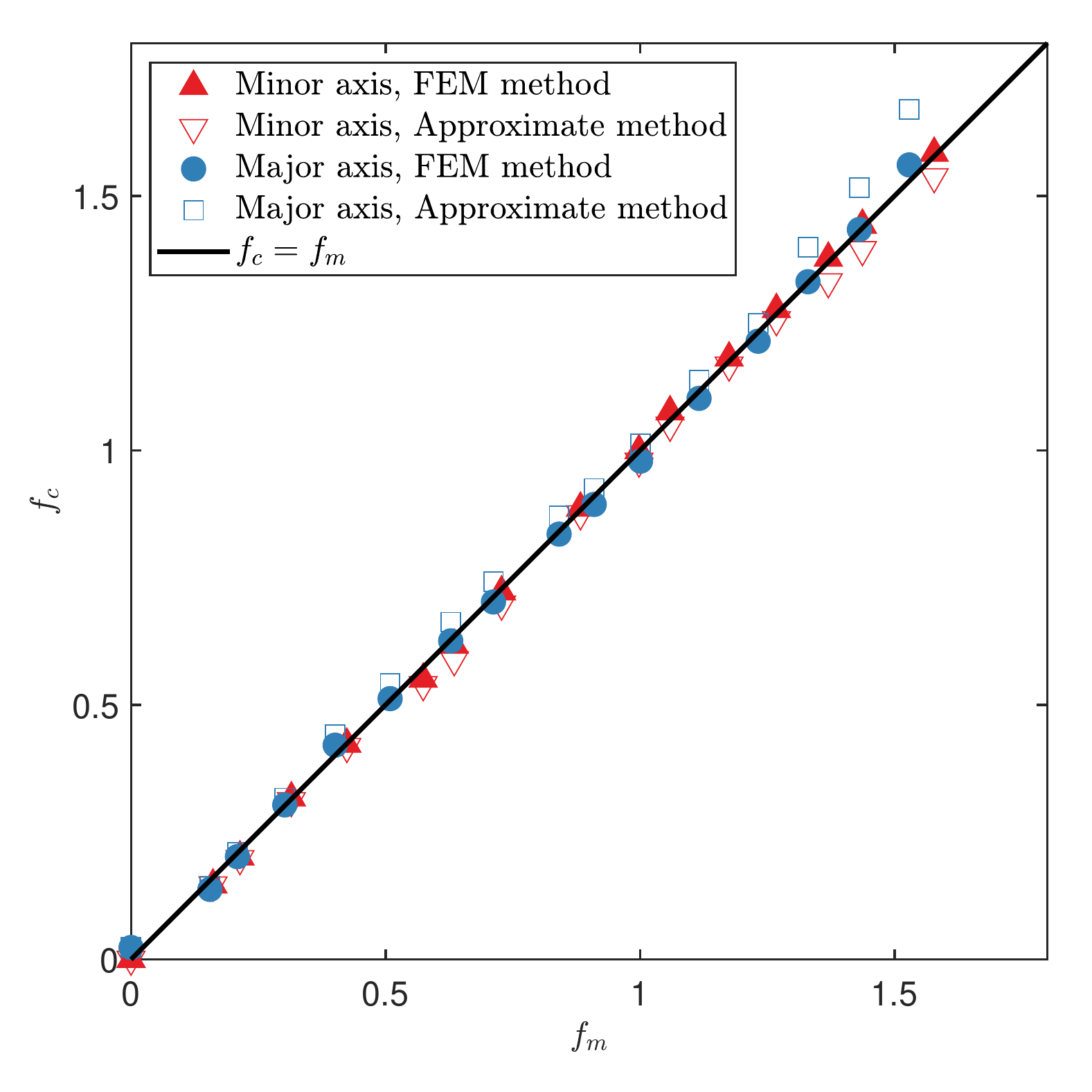} 
  \caption {Calculated contact forces $f_c$ versus measured contact forces $f_m$.}
 \label{fig:fm_fc}
\end{figure}

To test the accuracy of these two methods, we apply the diametric loading on the major axis and minor axis of an ellipse. We find these two methods both give reasonably good measurements of vector contact forces of the ellipse. The experimental stress images and the computed stress images subject to a series of loading forces are generated by the two methods as shown in Supplementary Fig. \ref{fig:ellipse_force_image}. More quantitatively, we plot the calculated contact forces $f_c$ by the two methods versus the contact forces $f_m$ measured by the force gauge in Supplementary Fig. \ref{fig:fm_fc}: the relative error is less than 5\% for typical force magnitude. Since the "approximate method" is almost 100 times faster than the "FEM method", the results in this Letter are given by the "approximate method". 

In principle, our revised force-inverse algorithm, works well in ellipses, is in principle readily applicable to other 2d particles of arbitrary shapes \cite{WangThesis}.

\section{Contact force distributions}

\begin{figure}
	\centerline{\includegraphics[width = 14.8 cm]{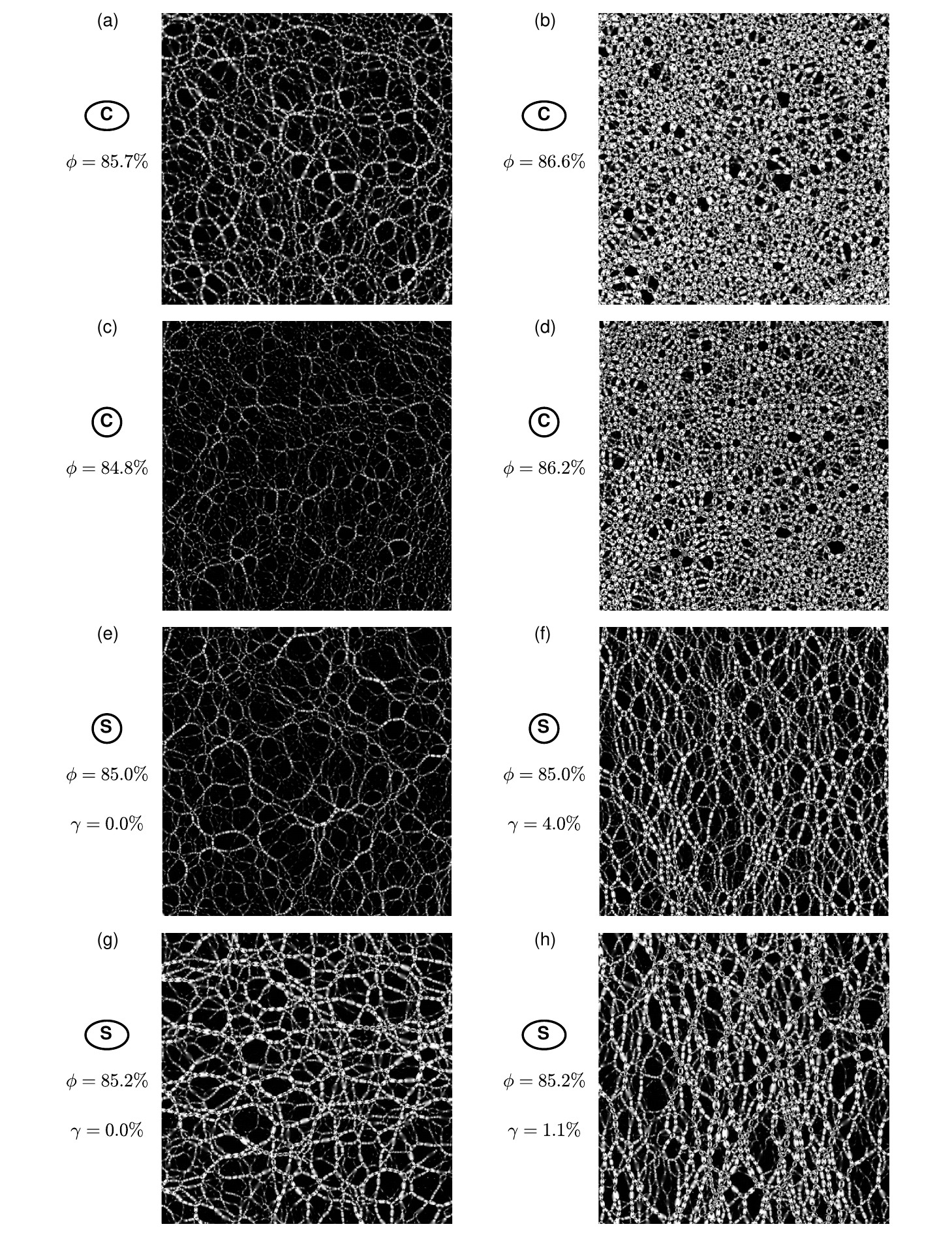}}
	\caption{Typical stress images of eight jammed states.
	In the legend, the circle/ellipse markers denote the experiments of circular/elliptical photoelastic particles, and the characters '\textbf{C}' and '\textbf{S}' within the markers denote the different experimental protocols of isotropic compression, i.e. '\textbf{C}', and the volume-conserved pure shear, i.e. '\textbf{S}'. The symbol $\phi$ denotes the packing fraction and $\gamma$ denotes the shear strain.}
	\label{fig:SFigure1}
\end{figure}

\begin{figure}
	\centerline{\includegraphics[width = 18.3 cm]{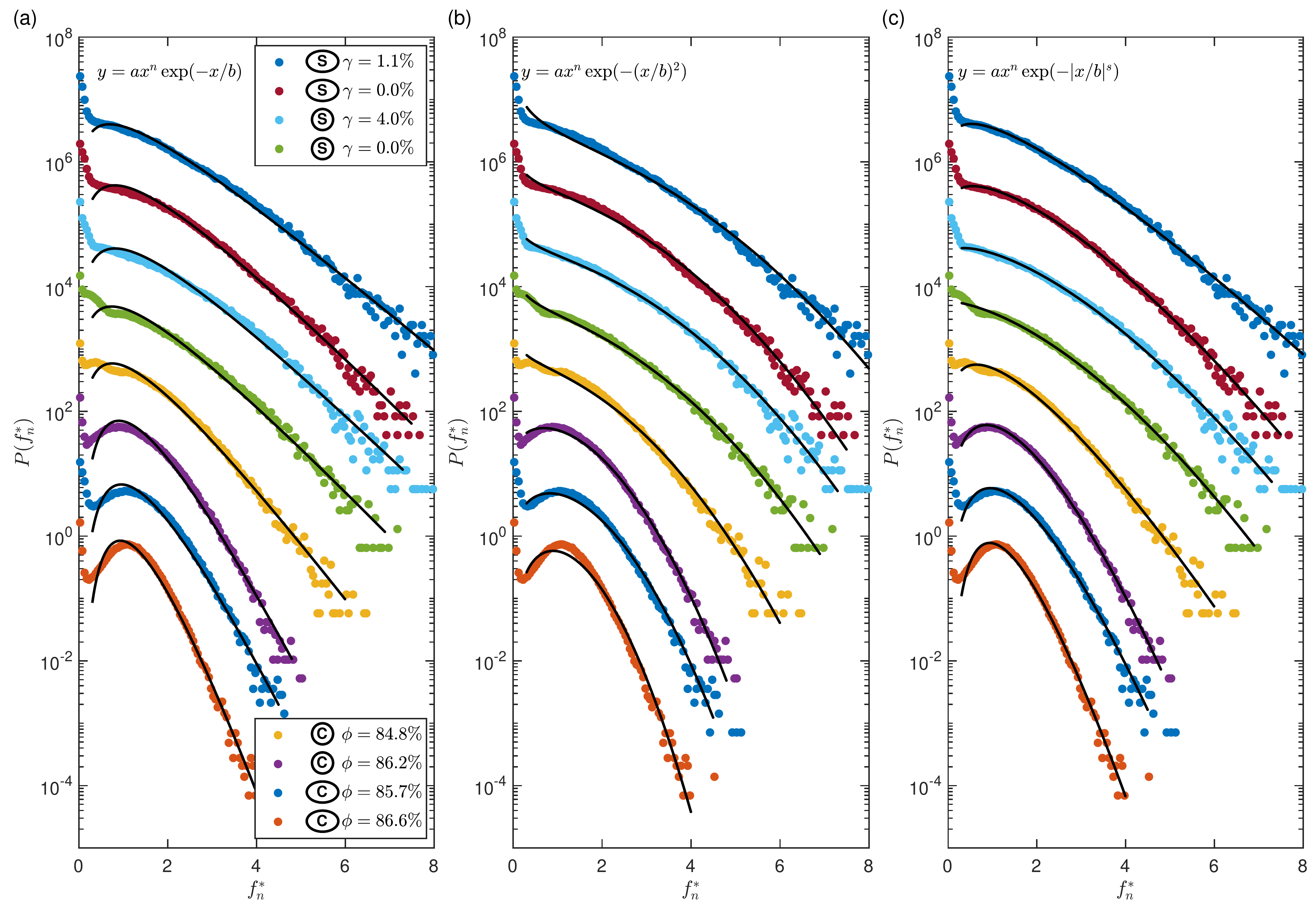}}
	\caption{
	\textbf{a}, The exponential fitting equation $y=ax^n\exp(-x/b)$. \textbf{b}, The Gaussian fitting equation $y=ax^n\exp(-(x/b)^2)$. \textbf{c}, The stretched exponential fitting equation $y=ax^n\exp(-|x/b|^s)$.
	Only data of $f_n^* > 0.3$ are fit, and the fitting parameters are listed in Supplementary Table \ref{table}. 
	In the legend, the circle/ellipse markers denote the experiments of circular/elliptical photoelastic particles, and the characters '\textbf{C}' and '\textbf{S}' within the markers denote the different experimental protocols of isotropic compression, i.e. '\textbf{C}', and the volume-conserved pure shear, i.e. '\textbf{S}'. The symbol $\phi$ denotes the packing fraction and $\gamma$ denotes the shear strain. In the shear experiments, the packing fraction $\phi = 85.0\%$ for disk packings and $\phi = 85.2\%$ for ellipse packings.}
	\label{fig:SFigure2}
\end{figure}

\begin{sidewaystable}
\begin{center}
\begin{tabular}{p{1.8cm} p{2.5cm} p{1.5cm} p{1.5cm} p{1.3cm} p{1.3cm} p{1.3cm} p{1.3cm} p{1.3cm} p{1.3cm} p{1.3cm} p{1.3cm} p{1.3cm} p{1.3cm}} 
\toprule
\multirow{2}{1.8cm}{Particle shape} & \multirow{2}{2.5cm}{Preparation protocol} & \multirow{2}{1.5cm}{Packing fraction} & \multirow{2}{1.5cm}{Shear strain}
 & \multicolumn{3}{l}{$y=ax^n\exp(-x/b)$} 
 & \multicolumn{3}{l}{$y=ax^n\exp(-(x/b)^2)$}  
 & \multicolumn{4}{l}{$y=ax^n\exp(-|x/b|^s)$} \\
 &  &  & 
 & a & n & b & a & n & b & a & n & b & s \\
\colrule
Ellipse    & Compression   & 86.6\% & 0.0\% & 188.89 & 5.05 & 0.18 & 1.22 & 1.23 & 1.15 & 23.34 & 3.79 & 0.36 & 1.20
\\ 
Ellipse    & Compression   & 85.7\% & 0.0\% & 34.21 & 3.74 & 0.25 & 0.76 & 0.66 & 1.44 & 3.29 & 2.16 & 0.67 & 1.35
\\ 
Disk       & Compression   & 86.2\% & 0.0\% & 27.21 & 3.30 & 0.27 & 0.78 & 0.43 & 1.49 & 2.08 & 1.49 & 0.85 & 1.47
\\ 
Disk       & Compression   & 84.8\% & 0.0\% & 5.67 & 1.75 & 0.42 & 0.53 & -0.36 & 2.02 & 1.50 & 0.80 & 0.92 & 1.30
\\ 
Disk       & Shear         & 85.0\% & 0.0\% & 2.81 & 1.35 & 0.54 & 0.42 & -0.47 & 2.42 & 0.57 & -0.01 & 1.76 & 1.61
\\ 
Disk       & Shear         & 85.0\% & 4.0\% & 2.24 & 1.41 & 0.58 & 0.37 & -0.40 & 2.57 & 0.52 & 0.13 & 1.76 & 1.56
\\ 
Ellipse    & Shear         & 85.2\% & 0.0\% & 2.44 & 1.45 & 0.56 & 0.38 & -0.43 & 2.53 & 0.66 & 0.37 & 1.42 & 1.41
\\ 
Ellipse    & Shear         & 85.2\% & 1.1\% & 1.62 & 1.02 & 0.67 & 0.34 & -0.68 & 2.93 & 0.81 & 0.47 & 1.15 & 1.20  \\ 
\botrule
\end{tabular}
\end{center}
\caption{Fitting parameters of $P(f_n^*)$ for the eight jammed states.}
\label{table}
\end{sidewaystable}

\begin{figure}
	\centerline{\includegraphics[width = 8.6 cm]{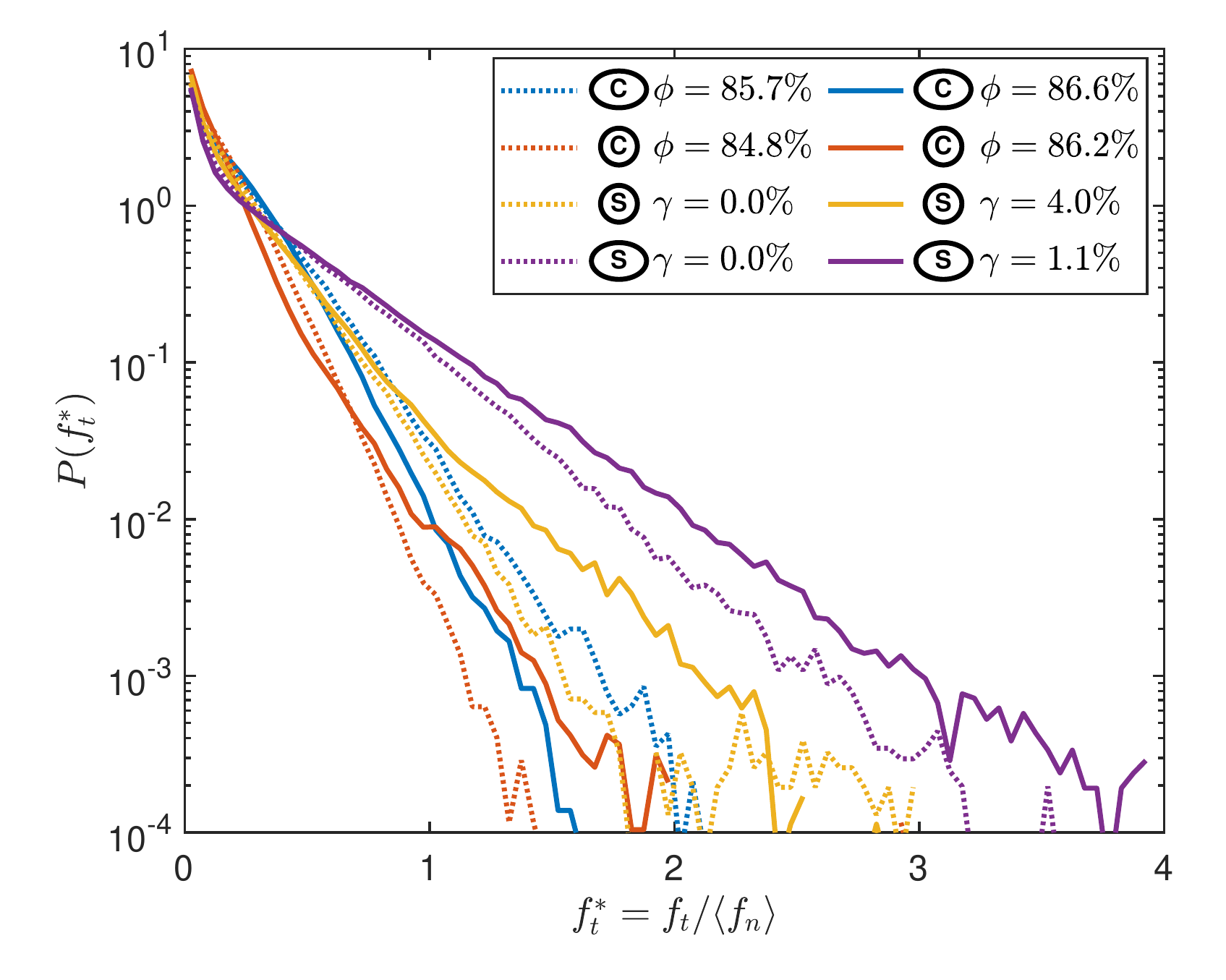}}
	\caption{Probability density functions of the normalized tangential forces $f_t^*=f_t/\langle f_n \rangle$. Here
	$\langle f_n \rangle$ refers to the mean normal force. In the legend, the circle/ellipse markers denote the experiments of circular/elliptical photoelastic particles, and the characters '\textbf{C}' and '\textbf{S}' within the markers denote the different experimental protocols of isotropic compression, i.e. '\textbf{C}', and the volume-conserved pure shear, i.e. '\textbf{S}'. The symbol $\phi$ denotes the packing fraction and $\gamma$ denotes the shear strain. In shear experiments, the packing fraction $\phi = 85.0\%$ for disk packings and $\phi = 85.2\%$ for ellipse packings.}
	\label{fig:SFigure3}
\end{figure}

\begin{figure}
	\centerline{\includegraphics[width = 14.8 cm]{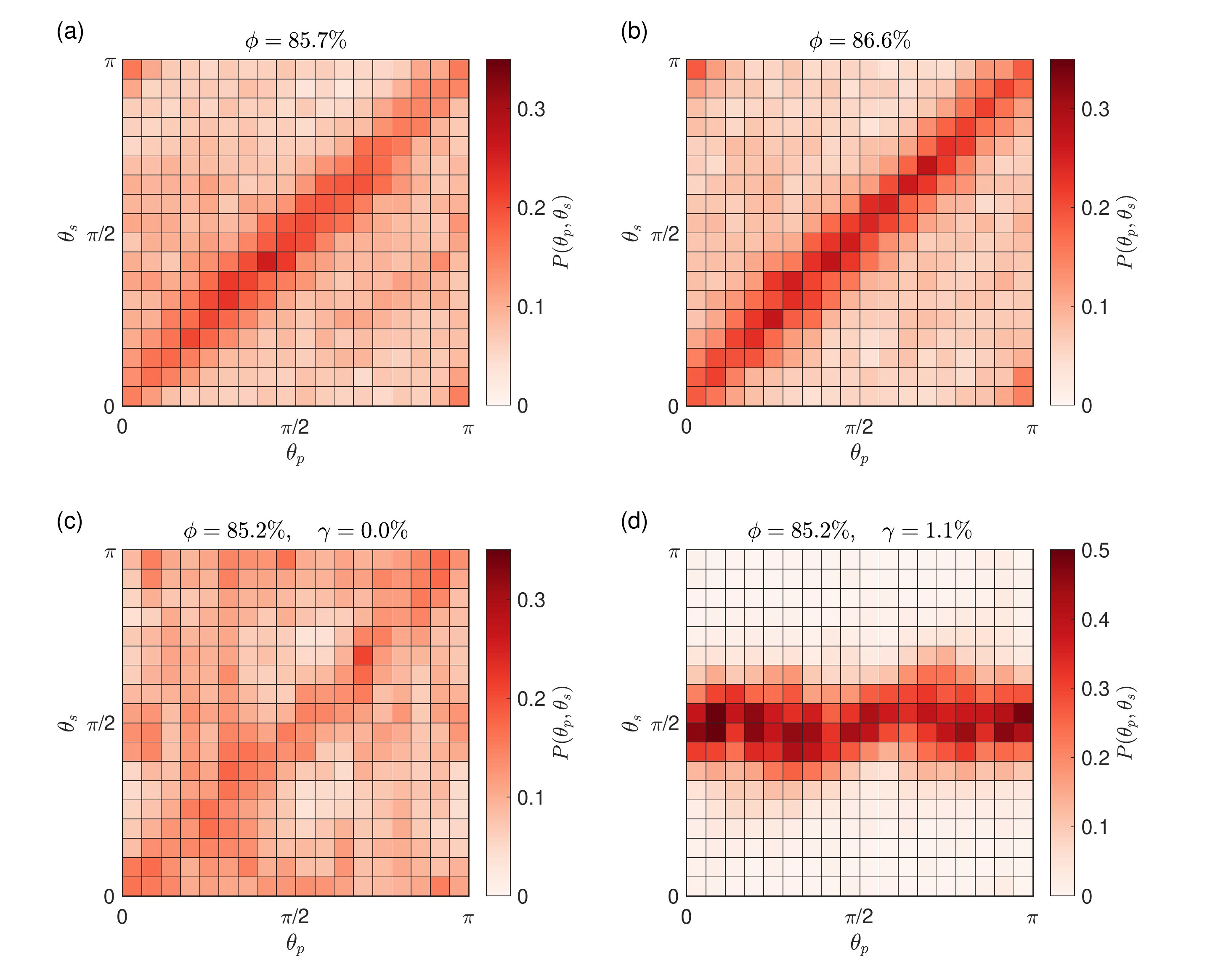}}
	\caption{Joint probability density functions $P(\theta_p,\theta_s)$ of the particle orientations $\theta_p$ and the stress orientations $\theta_s$ for ellipse packings.
	$P(\theta_p,\theta_s)$ of ellipse packings subject to isotropic compression with the packing fraction of \textbf{a}, $\phi=85.7\%$, and \textbf{b}, $\phi=86.6\%$. $P(\theta_p,\theta_s)$ of ellipse packings subject to pure shear at the strain of \textbf{c}, $\gamma=0.0\%$, and \textbf{d}, $\gamma=1.1\%$.}
	\label{fig:SFigure4}
\end{figure}

\begin{figure}
	\centerline{\includegraphics[width = 6 cm]{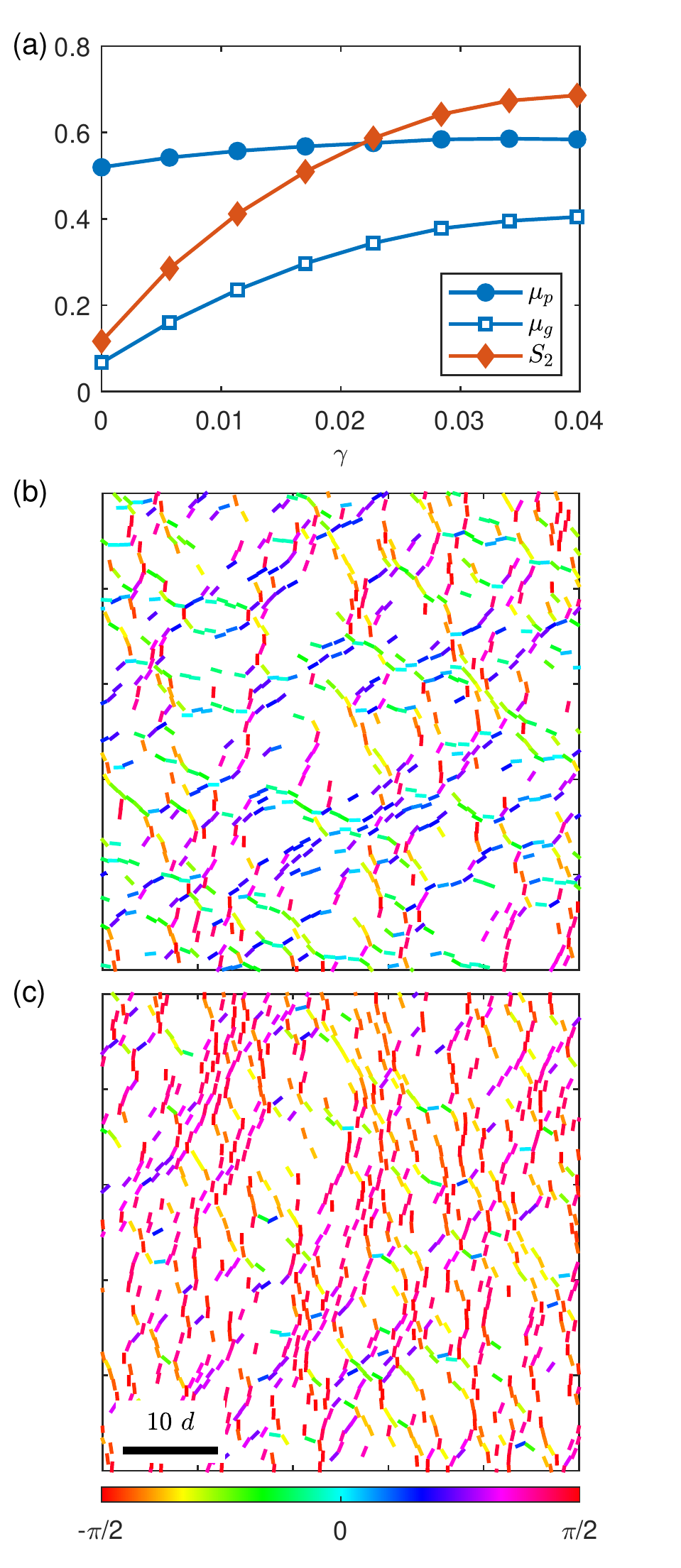}}
	\caption{
	\textbf{a}, The global stress ratios $\mu_g$, the particle stress ratios $\mu_p$ and the nematic order of particle principle stress orientations $S_2$ as a function of shear strain $\gamma$. The spatial distributions of the particle principle stress orientations in \textbf{b}, where $\gamma=0$, and in \textbf{c}, where $\gamma=0.04$. The painted bars denote the particle principle stress orientations. Only particles with pressure larger than the mean value are shown.}
	\label{fig:SFigure5}
\end{figure}

\newpage
\
\newpage
\
\newpage
\
\newpage
\
\newpage